Biomechanical Classification of Judo Throwing Techniques (Nage Waza)

Attilio Sacripanti

ENEA
University of Tor Vergata  Roma
Italian Wrestling. Weightlifting and Judo Federation.

Abstract
In this paper it is applied the classic mechanical  point of view to classify all the movements known as throwing judo techniques; the application of the Newtonian physical methods and principles is able to rationalize the whole matter and to group under only two very simple and clear principles all the movements , before grouped in many generic and not very clear way.

I) HISTORICAL SURVEY

The classification of standard Judo throwing techniques (Nage Waza) was born from the following didactic requirement - To group the standard techniques under logical criteria for an easier understanding and useful systematic study.

The two problems of Classification and Teaching arrangement were tackled and solved by dr. Kano (founder of Judo) and his assistants in a scientific way, according to the knowledge of their time.

The first classification of Nage Waza (1882) was carried out by a real proto-biomechanical method.

As a matter of fact, the standard techniques were classified by parts of Tori's (the attacker) body which work as greater contact-point for energy transfer in throwing.

From that we have the "Kodokan Classification": Te waza = shoulder, arm and hand techniques; Koshi waza = hip techniques; Ashi waza = leg techniques; Sutemi waza = body-abandoning techniques or sacrifice techniques (Tab. 1).

This admirable classification is simple, intelligible and almost perfect.

That is the reason of its long success even if it has some shades in itself; in fact it is unusual to see hands, hips or legs of Tori working alone in throwing.

Besides the Kodokan classification uses a different way to classify the body-abandoning techniques, which are classified by the body side touching the mat: Ma sutemi waza and Yoko sutemi waza, body abandoning techniques on his own back and on his own side.

In his golden book  My study of J u d o , G . Koizumi classified the standard judo techniques by Uke's (the defender) body motion.

This study, performed as systematical analysis of Nage Waza, permits to group the standard techniques in three sets, according to the basis of the technical principles.

Namely the "Koizumi Classification": Kuruma waza = wheel techniques.

Throws which are effected in such a way, that Uke's body is curled and turned as a wheel.

Tenbin waza = Scale techniques. Throws which are effected by tripping or propping Uke's body, as a scale while it is held as a pole.

Tsumazukase waza = Tripping techniques. Throws which are effected by tripping Uke's foot or leg, preventing it from moving to regain or maintain stability.

In recent years A. Geesink and G. R. Gleeson have synthesized other kinds of classification.

The former Duch champion, in his work, emphasizes the dynamic role played by biodynamical chains, while the latter, English senior coach surely the subtlest student of western judo ones, in his historical book "Judo for the West", makes a very clever classification based on his enormous theoretical and empirical knowledge.

This classification collects the throwing techniques under two groups; First-Class of "turning" Uke's body around an obstacle (hip, leg, etc.) - Second- Class of "striking" Uke 's legs.
This historical survey through various classifications introduces the attempt to rationalize the matters in a scientific way, looking for the basic physical principles of Nage Waza.

## TABLE 1
### FIRST KODOKAN CLASSIFICATION (1885)

| TE WAZA | KOSHI WAZA | ASHI WAZA |
|---|---|---|
| Uki waza | Uki goshi | Okuri ashi harai |
| Seoi nage | Harai goshi | Sasae tsurikomi ashi |
| Kata guruma | Tsuri komi goshi | Uchi mata |
| Tai otoshi | Koshi guruma | Hiza guruma |
| Obi otoshi | O goshi | O soto gari |
| Seoi otoshi | Ushiro goshi | De ashi harai |
| Uki otoshi | Hane goshi | Ko uchi gari |
|  | Tsuri goshi | Ko soto gari |
|  | Utsuri goshi | Harai tsurikomi ashi |
|  |  | O uchi gari |
|  |  | Yama arashi |
|  |  | O soto guruma |
|  |  | O soto otoshi |

| MA SUTEMI WAZA | YOKO SUTEMI WAZA | |
|---|---|---|
| Tomoe nage | Yoko gake | Uchi makikomi |
| Ura nage | Yoko guruma | Tani otoshi |
| Sumi kaeshi | Yoko otoshi | |
| Hikkikomi gaeshi | Daki wakare | |
| Tsuri otoshi | Yoko wakare | |
| Tawara gaeshi | Soto makikomi | |

II) BIOMECHANICAL CLASSIFICATION
A biomechanical analysis of judo throwing techniques must be dealt with in the following steps: firstly by simplification and secondly by generalization.
As a simplification principle for the problem of classes of forces involved, at first we can use the differential method pointed out by dr. Kano: subdivision of the throwing movement in three steps:
 1° Tsukuri (preparatory movements aimed at throwing out of balance Uke's body);
 2° Kuzushi (the final unbalancing action); and
 3° Kake (execution of movements aimed at throwing), and later we analyse the motion of Uke's body cutting out secondary forces.
Then we generalize the classes of forces, putting out the inner physical principles of standard judo throwing techniques.
This method, applied to Nage Waza. is able to group 77 throwing techniques (40 Kodokan go kyo and 37 others) under two dynamic principles only.
It is correct to remark that this is one of many possible biomechanical classifications, and we select it for its valuable simplicity and immediateness.

It comes very handy to find the «General Principles", first to define two corollaries on direction of forces (Static Analysis), and then to analyze Uke's body flight paths (Dynamical Analysis) and their symmetries.

STATIC ANALY SIS
Principium of resolution of forces.
These two corollaries determine the whole directional problem of static use of forces to execute throws.

UNBALANCES
1) Forces are effective and can be applied, on the horizontal plane, on the whole round angle (360°).

Unified under these terms are the biomechanical problems of forces employed for unbalancing Uke's body. (Tsukuri-Kuzushi steps).

THROWS
2) Forces are effective and can be applied, on the vertical plane, nearly for the width of a right angle (90°).

Determined under these terms are the biomechanical problems of forces employed for throwing Uke's body. (Kake step). Real limits of throwing forces can be obtained with an angle of nearly 45 degrees, up or down a horizontal line, because the resistance caused by Uke's body structure or by force of gravity, beyond these angles, allows throwing again but with more waste of energy.

DYNAMICAL ANALYSIS
Principium of composition of forces  - Study of flight paths and its symmetries

If in the space the composition of forces obeys, at the same time, the previous static two corollaries the solution of dynamical problem (considering time) goes through the study of flight paths and their symmetries.
Trajectories along which Uke's  body moves during flight following throwing can be collected under two simple types or their composition:   Circular paths and Helicoidal paths.

I) CIRCULAR PATH: SPHERICAL SYMMETRY
For throwing techniques in which the limbs of Uke's body follow a circular path

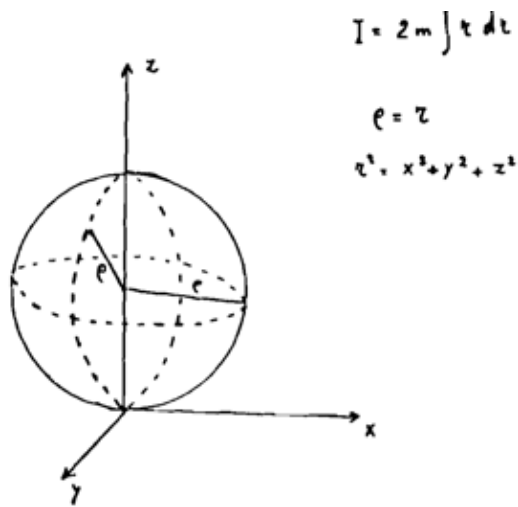

$$I = 2m \int r \, dr$$

$$\rho = r$$

$$r^2 = x^2 + y^2 + z^2$$

the radius of circumference coincides with the distance r from the rotation axis of inertial momentum. These techniques have spherical symmetry and circumference that is the "geodetic line" of sphere (the shortest line between two points) is the path of minimum work then the trajectory of least waste of energy which extreme parts of Uke's body can cover.

II) HELICOIDAL PATH: CYLINDRICAL SYMMETRY
For throwing techniques in which Uke's body follows a helicoidally path the bending radius of helix is proportional to distance r from the rotation axis of inertial momentum.
These techniques have cylindrical symmetry and the helix that is the «geodetic line» of right cylinder is also the path of minimum work or the trajectory of least waste of energy which Uke's body can cover.

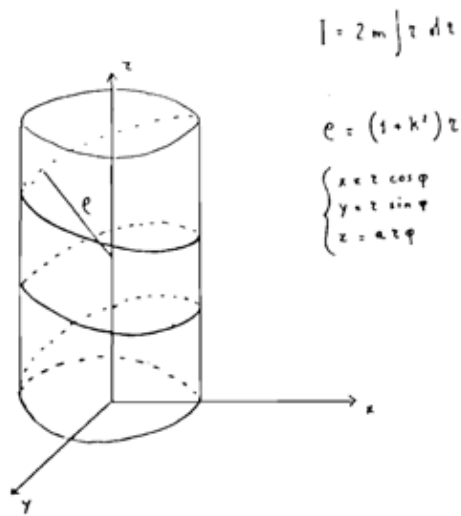

If we think both ,of the two corollaries on direction of forces (Static Analysis) and of the study of trajectories followed by Uke's body (Dynamical Analysis) it is possible to state the two dynamical principles of biomechanical classification, which show the inner mechanisms of throwing techniques.

A) Techniques where Tori makes use of a couple of forces for throwing Uke.
B) Techniques where Tori makes use of physical lever for throwing Uke.

Movements, that seemingly make different throwing techniques, in appearance but non in biomechanical essence, can be collected in Tsukuri - Kuzushi stages and other preparatory actions (Taisabaki, etc.).
We think this classification, grounded on clear scientific criteria, is very suitable to give an easier understanding of physical principles linking judo throwing techniques.

A) Techniques of couple of forces.
In the first group, we found all throws produced by sweeping away legs and pulling or pushing Uke's body in the opposite direction simultaneously.
The techniques of "Group of couple of forces" can be classified by parts of Tori's body which apply the couple of forces on Uke's body.

Namely: two arms, arm and leg, trunk and leg, trunk and arms, two legs. (Tab. 2).

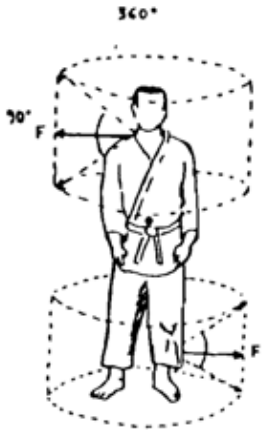
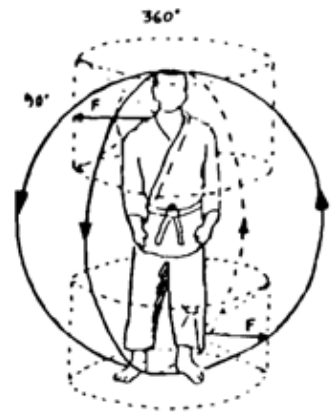

GROUP OF COUPLE OF FORCES

STATICAL CONDITIONS
Complete unbalance angle 360°
Right throwing angle 90°

DYNAMICAL CONDITIONS
Circular flight path
Spherical symmetry

This biomechanical classification is able to show new likeness not evident in standard techniques, e.g. the asymmetry face-back of human body explains astonishing examples of biomechanical likeness: O Soto Gari and Uchi Mata are the same techniques applying the couple of forces on Uke's body.
It is very interesting to note that most throws of couple of forces (the ones applied by Tori standing on one leg) can be led to only one Tori's basic action: rotation on coxo-femoral articulation with three degrees of freedom each of it, lying into the three symmetry planes of the human body.
First: rotation of trunk-leg set on coxo-femoral articulation around a horizontal lateral-lateral axis of rotation. ( motion lying in the sagittal plane )

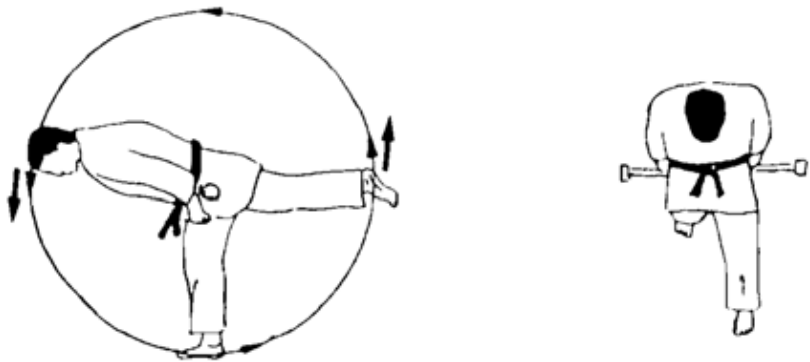

Second: rotation of trunk-leg set on coxo-femoral articulation around a horizontal antero-posterior axis of rotation. ( motion lying in the frontal plane )

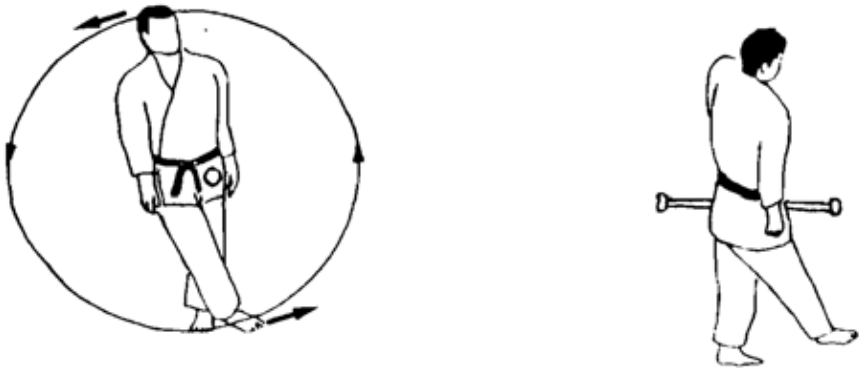

Third: rotation of trunk-leg set on coxo-femoral articulation around a vertical axis of rotation. (motion lying in the transverse plane)

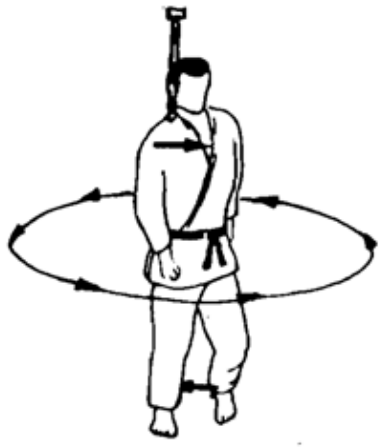

This points out the fundamental role played by COXO-FEMORAL ARTICULATION in this group, and entails that this athlete's articulation must be provided with a great flexibility.

B) Techniques of physical lever
 In the second group we found all throws produced by turning Uke's body round a stopping point (hip, leg, foot, etc.). The techniques of "Group of physical lever" can be classified by length of arm of lever, applied on Uke's body.
Namely: minimum arm (fulcrum under Uke's waist), medium arm (fulcrum under Uke's knees), maximum arm (fulcrum under Uke's malleola), variable arm (variable fulcrum from the waist down to Uke's knees) (Tabl. 3).
Because in this group throws of "minimum arm" are energetically unfavorable (greatest force applied), that clears why, for competition, people like more to turn them in throws of variable arm, pulling down fulcrum under Uke's waist more and more. That means less waste of energy.

## GROUP OF MOMENTUM OF FORCE

GROUP OF PHYSICAL LEVER
o
GROUP OF MOMENTUM OF FORCE

DYNAMICAL CONDITIONS
Flight path: minimum and medium arm Helicoidal
             maximum arm Circular or Helicoidal
Symmetry   : minimum and medium arm Cylindrical
             maximum arm Spherical or Cylindrical

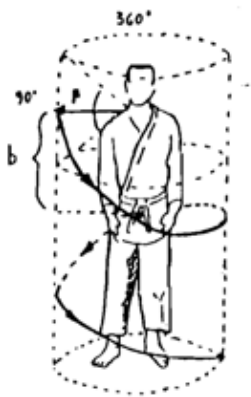

Minimum arm
Maximum applied force

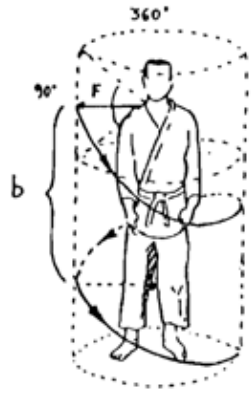

Medium arm
Medium applied force

Fulcrum of physical lever

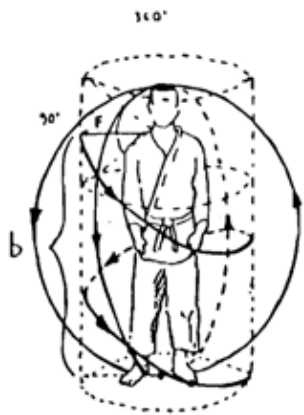

Maximum arm
Minimum applied force

Again this biomechanical classification shows, in this group, likeness examples: classical Ashi Guruma and Hiza Guruma are the same techniques applying a lever of medium arm on Uke's body.
It is very interesting to note that most throws of group of lever can be led to only one Tori's basic action, if we do not consider the several Tori's legs positions: rotation of trunk on waist around a generic variable axis of rotation.

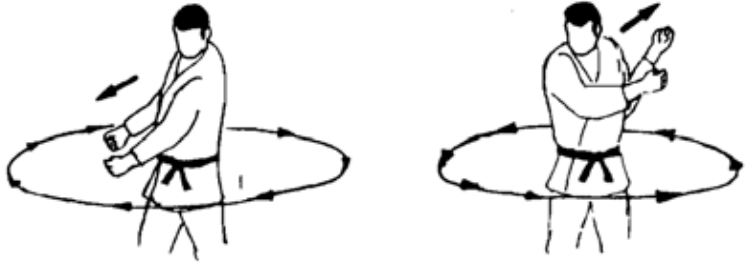

Also the body-abandoning techniques must be classified as throws of "Group of physical lever" with maximum arm, in this case the stopping point (fulcrum) is given by friction between foot and mat (tatami).
Although they are more favorable energetically speaking, the starting force (body weight falling down) is applied with an angle greater of 45° (see Il corollary of static analysis).
That is matters need high, directional help by arms or legs, for rightly throwing Uke's body.
At the light of our analysis we think the clarification of basic physical principles and the evidence of basic action, proving the leading role of Tori's pelvic belt, can be useful for a better understanding of inner mechanisms and for looking at new improvements in training theory, which should prevent erroneous methods and possible damages in the joints.

# TABLE 2

## BIOMECHANICAL CLASSIFICATION

## TECHNIQUES OF COUPLE OF FORCES
Couple Applied by

| ARMS | ARM/S AND LEG | | |
|---|---|---|---|
| Kuchiki daoshi | De ashi barai | Ko uchi gari | Yoko gake |
| Kibisu gaeshi | Okuri ashi barai | Ko soto gari | Ko soto gari |
| Kakato gaeeshi | Ko uchi barai | O uchi gari | O uchi gake |
| Te guruma | O uchi barai | Ko uchi gake | Ko soto gake |
| | Tsubame gaeshi | Harai tsukiromi ashi | |

| TRUNK AND ARMS | | TRUNK AND LEG | |
|---|---|---|---|
| Morote gari | O soto gari | Uchi mata | Harai goshi |
| | O soto guruma | Hane goshi | Yama arashi |
| LEGS | O soto otoshi | Hane makikomi | O uchi sutemi |
| | O tsubushi | Okurikomi uchimata | |
| Kani basami | | | |

# TABLE 3

## BIOMECHANICAL CLASSIFICATION

## TECHNIQUES OF PHYSICAL LEVER
Lever applied with

### MINIMUM ARM (FULCRUM UNDER UKE'S WAIST)

| | | |
|---|---|---|
| O guruma | Sukui nage | Tawara gaeshi |
| Kata guruma | Ushiro goshi | Ura nage |
| Tama guruma | Utsuri goshi | Ganseki otoshi |
| Obi otoshi | Soto makikomi | Uchi makikomi |

### MEDIUM ARM (FULCRUM UNDER UKE'S KNEES)

Ashi guruma    Hiza guruma

## MAXIMUM ARM (FULCRUM UNDER UKE'S MALLEOLA)

| | | |
|---|---|---|
| Uki otoshi | Uki waza | Sasae tsurikomi ashi |
| Yoko otoshi | Dai sharin | Seoi otoshi |
| Sumi otoshi | Tomoe nage | Hiza seoi |
| Ura otoshi | Sumi gaeshi | Suwari seoi |
| Waki otoshi | Hikkomi gaeshi | Obi seoi |
| Tani otoshi | Yoko guruma | Suso seoi |
| Tai otoshi | | |

## VARIABLE ARM (VARIABLE FULCRUM FROM THE WAIST DOWN TO KNEES)

| | | |
|---|---|---|
| Tsuri komi goshi | Uki goshi | Kata seoi |
| Sasae tsurikomi goshi | O goshi | Seoi nage |
| Ko tsuri komi goshi | Koshi guruma | Eri seoi nage |
| O tsurikomi goshi | Kubi nage | Morote seoi nage |
| Sode tsurikomi goshi | | |

III) JUDO SKILLS AND THEIR TREND IN TOP COMPETITION

With the continuous evolution of competitive style, for a right training of «top» athletes, we need a better understanding of judo skills and their trend in top competitions.

Obviously the basic physical principles of standard judo techniques and judo skills (competitive throwing techniques) are the same, but the dynamical conditions are quite different (fast movements, more opposition, timing, change of speed, etc.).

The right way to learn the connection with standard techniques, and to understand the competitive evolution of judo skills, was signed for the first time, by dr. Kano with the formulation of principium of "Maximum effect with minimum effort".

This principium can be translated in two useful biomechanical remarks:

1) To improve techniques to win, signifies to produce maximum economy of movements.

2) To better techniques to win, signifies to produce maximum economy of strain.

The explication of remarks is: in competition champions, consciously or unawares, tend to find the right methods of minimizing the total energy to win.

In two ways: either minimizing his own muscular strain with right changes in throwing patterns (e.g. pulling down the fulcrum), or minimizing the defensive antagonist capability with right changes in attack directions (e.g. attack in the direction of the adversary side where the resistive muscular capability is weaker).

These are two biomechanical rules, which give us a better understanding of trend of judo skills in top competitions.